\documentclass[reprint,amssymb,aps,prl,superscriptaddress]{revtex4-2}

\usepackage{CJK}
 \usepackage{xcolor}
\usepackage{amsfonts,amssymb,amsmath,amsthm,mathtools,wasysym,bm}
\usepackage{breqn}
 \usepackage{ulem}
\makeatletter
\let\cat@comma@active\@empty
\makeatother
\usepackage{physics}
\usepackage{graphicx}
\usepackage{xcolor}
\usepackage[caption=false]{subfig}
\newcommand{\phantomsubcaption}[1]{
    {
        \captionsetup[subfloat]{farskip=0pt,captionskip=0pt}
        \captionsetup[subfigure]{labelformat=empty}
        \subfloat{#1}
    }%
}
\usepackage{tikz}
\usepackage{hyperref}
\hypersetup{
    colorlinks=true,
    linkcolor=blue,
    citecolor=green,
    urlcolor=cyan,
    }
\usepackage{cleveref}
\usepackage[inline]{enumitem}
\usepackage{diagbox}

\usepackage{pdfpages}
\makeatletter
\AtBeginDocument{\let\LS@rot\@undefined}
\makeatother

\begin{document}

\title{
Cavity-Vacuum-Induced Chiral Spin Liquids in Kagome Lattices: Tuning and Probing Topological Quantum Phases via Cavity Quantum Electrodynamics
}
\author{Chenan Wei}
\affiliation{Department of Physics, University of Massachusetts, Amherst, Massachusetts 01003, USA}
\affiliation{A. Alikhanyan National Science Laboratory, Br. Alikhanian 2, Yerevan 0036, Armenia}

\author{Liu Yang}
\affiliation{Tsung-Dao Lee Institute, Shanghai Jiao Tong University, Shanghai 200240, China}
\affiliation{School of Physics and Astronomy, Shanghai Jiao Tong University, Shanghai 200240, China}

\author{Qing-Dong Jiang}
\email{qingdong.jiang@sjtu.edu.cn}
\affiliation{Tsung-Dao Lee Institute, Shanghai Jiao Tong University, Shanghai 200240, China}
\affiliation{School of Physics and Astronomy, Shanghai Jiao Tong University, Shanghai 200240, China}
\affiliation{Shanghai Branch, Hefei National Laboratory, Shanghai 201315, China}

\date{\today}
\begin{abstract}
Topological phases in frustrated quantum magnetic systems have captivated researchers for decades, with the chiral spin liquid (CSL) standing out as one of the most compelling examples. Featured by long-range entanglement, topological order, and exotic fractional excitations, the CSL has inspired extensive exploration for practical realizations. In this work, we demonstrate that CSLs can emerge in a kagome lattice driven by vacuum quantum fluctuations over the non-interacting vacuum within a single-mode gyrotropic cavity. The gyrotropic cavity imprints quantum fluctuations with time-reversal symmetry breaking and stabilizes a robust CSL phase without external laser excitation.
Moreover, we identify experimentally accessible observables---such as average photon number and transport properties---that reveal connections between photon dynamics and the emergent chiral order. Our findings establish a novel pathway for creating,  controlling, and probing topological and symmetry-breaking quantum phases in strongly correlated systems.
\end{abstract}
\maketitle
\paragraph{Introduction.}

Quantum spin liquids (QSLs) are unique “quantum disordered” ground states where zero-point fluctuations prevent conventional magnetic order. A notable subset, chiral spin liquids (CSLs), break time-reversal symmetry and have been studied in strongly correlated systems for their potential to host topologically ordered states and anyonic excitations vital for quantum computing \cite{kalmeyer1987equivalence,kalmeyer1989theory,wen1989chiral,hu2016variational,huang2021quantum,zhang2021SU(4)}. Advances in theories of moat bands and spontaneous chiral breaking in fermionic systems \cite{sedrakyan2009fermionic,sedrakyan2012composite,sedrakyan2014absence,sedrakyan2015statistical,sedrakyan2015spontaneous,wang2018chern,maiti2019fermionization,wang2022emergent,wei2023chiral,wang2023excitonic,wei2024unveiling,jafarizadeh2025chiral} have deepened our understanding of CSL formation. Their chiral nature appears in responses to electromagnetic probes in both equilibrium and nonequilibrium conditions \cite{banerjee2023electromagnetic} and in the emergence of circulating spin currents, driven by geometric frustration (e.g., in kagome lattices) or external perturbations such as staggered magnetic fields and spin-orbit coupling.

One promising approach to induce chiral ordering is cavity quantum electrodynamics (cQED). In cQED, strong electron-photon interactions within an optical cavity can significantly change a material’s electronic, magnetic, topological, and localization properties \cite{kiffner2019manipulating,mendez2020renyi,schlawin2022cavity,mendez2023from,guo2024vacuum,bacciconi2024theory}. Coupling quantum materials to cavity modes enables dynamic tuning of the Hamiltonian, offering precise control over quantum many-body systems, including QSLs \cite{chiocchetta2021cavity}.

In a gyrotropic cavity, the engineered polarization of the electromagnetic field breaks time-reversal symmetry, providing a new route to stabilize topological chiral phases via quantum fluctuations \cite{espinosa2014semiconductor,sedov2022cavity,tokatly2021vacuum,schafer2018Ab,wang2019cavity,jiang2019quantum,baranov2020circular,baranov2020ultrastrong,owens2022chiral,bloch2022strongly,valagiannopoulos2022electromagnetic,zhu2024experimental,yang2024emergent,jiang2024engineering}.
A simple method to realize such a cavity uses a Faraday rotator (e.g., a ferromagnetic layer) together with high-quality metallic mirrors \cite{hubener2021engineering,voronin2022single,baranov2023toward,jarc2023cavity,vinas2023controlling}.
Pioneering works based on the Floquet method, such as using laser driving to induce a gauge flux for realizing QSL states \cite{quito2021floquet,sriram2022light,kumar2022floquet,sun2023engineering,hui2019flux,mambrini2024quantum}, have proven highly fruitful.
However, external electromagnetic driving can push the system out of equilibrium, potentially causing heating, loss of quantum coherence, and transient behavior. The vacuum cavity approach intrinsically avoids laser-induced heating and supports long-lived equilibrium quantum states. This inherent stability preserves quantum coherence and makes our proposal accessible to transport experiments.

\begin{figure}
    \centering
    \begin{minipage}{\linewidth}
    \includegraphics[width=\linewidth]{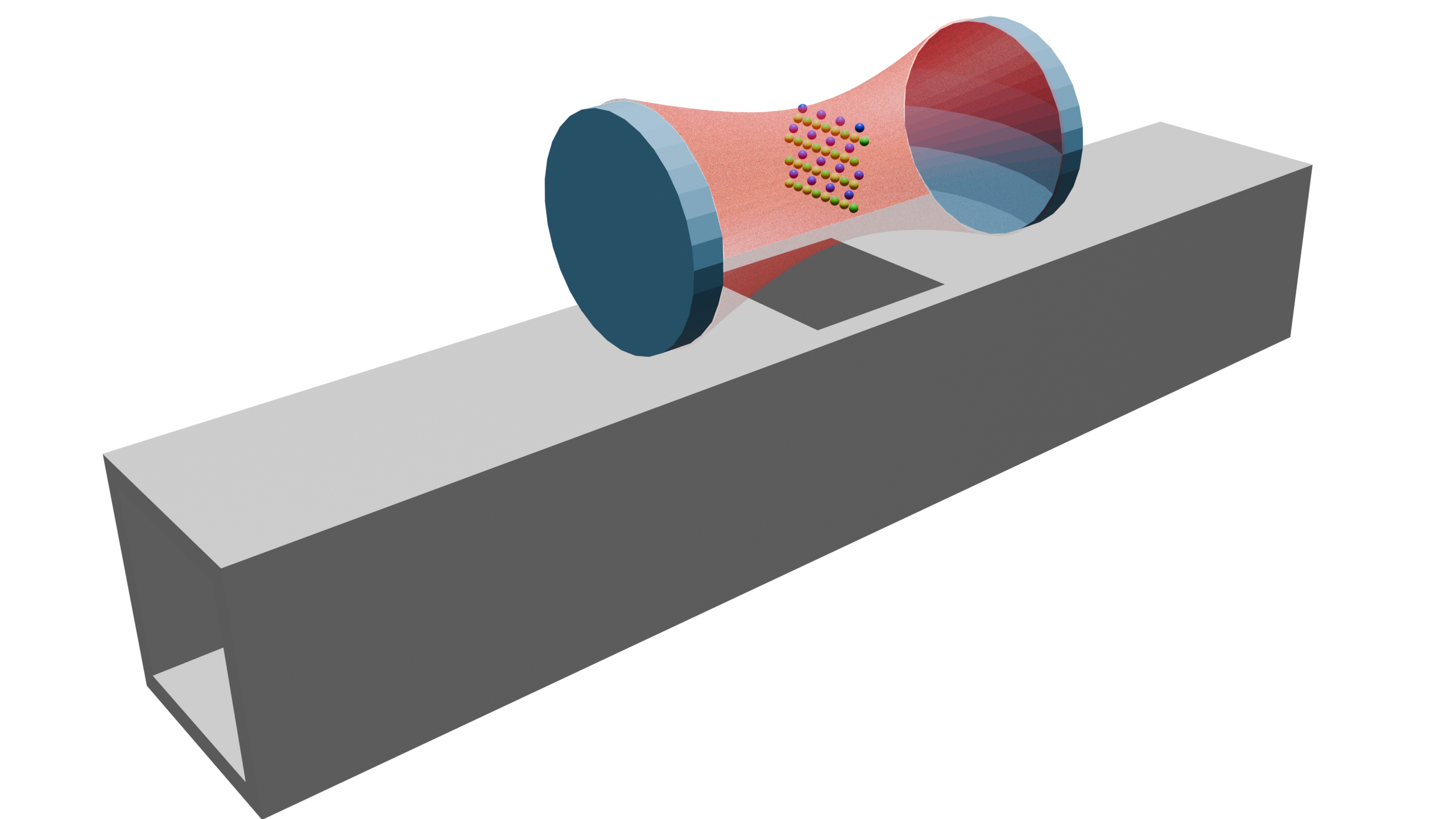}
    \begin{picture}(0,0)
    {\def\unitlength{}
        \put(-0.31\linewidth,0.3\linewidth){waveguide}
        \put(-0.33\linewidth,0.55\linewidth){gyrotropic cavity}
    }
    \end{picture}
    \end{minipage}
    \caption{\label{schematic}
    The schematic diagram of the setup. The kagome lattice is placed inside a gyrotropic cavity. The gyrotropic cavity is coupled to a waveguide, which is used for transport measurements.
}
\end{figure}

In this work, we explore a novel mechanism for inducing CSL phases by coupling a kagome lattice to a gyrotropic cavity (see \cref{schematic} for a schematic diagram of the setup). This setup leverages the interaction between virtual photons from vacuum fluctuations and electrons within the kagome lattice to dynamically break time-reversal symmetry and establish chiral order. The ability to control topological phases via a tunable cavity vacuum represents a remarkable alternative to prior studies on CSLs in strongly correlated systems.
Specifically, we investigate how coupling the lattice system to cavity modes can enhance quantum fluctuations and induce the chiral order needed for CSL phases. Furthermore, we propose experimentally accessible transport properties that link photon dynamics with the emergent chiral spin order.
By examining the interplay between the frustrated kagome geometry and the electromagnetic quantum fluctuations within the cavity, we aim to provide a new avenue for engineering and probing topologically nontrivial order in frustrated magnetic systems.

\paragraph{Kagome lattice coupled to a gyrotropic cavity.}

\begin{figure}
    \centering
    \begin{minipage}{0.15\textwidth}
    \includegraphics[width=\linewidth]{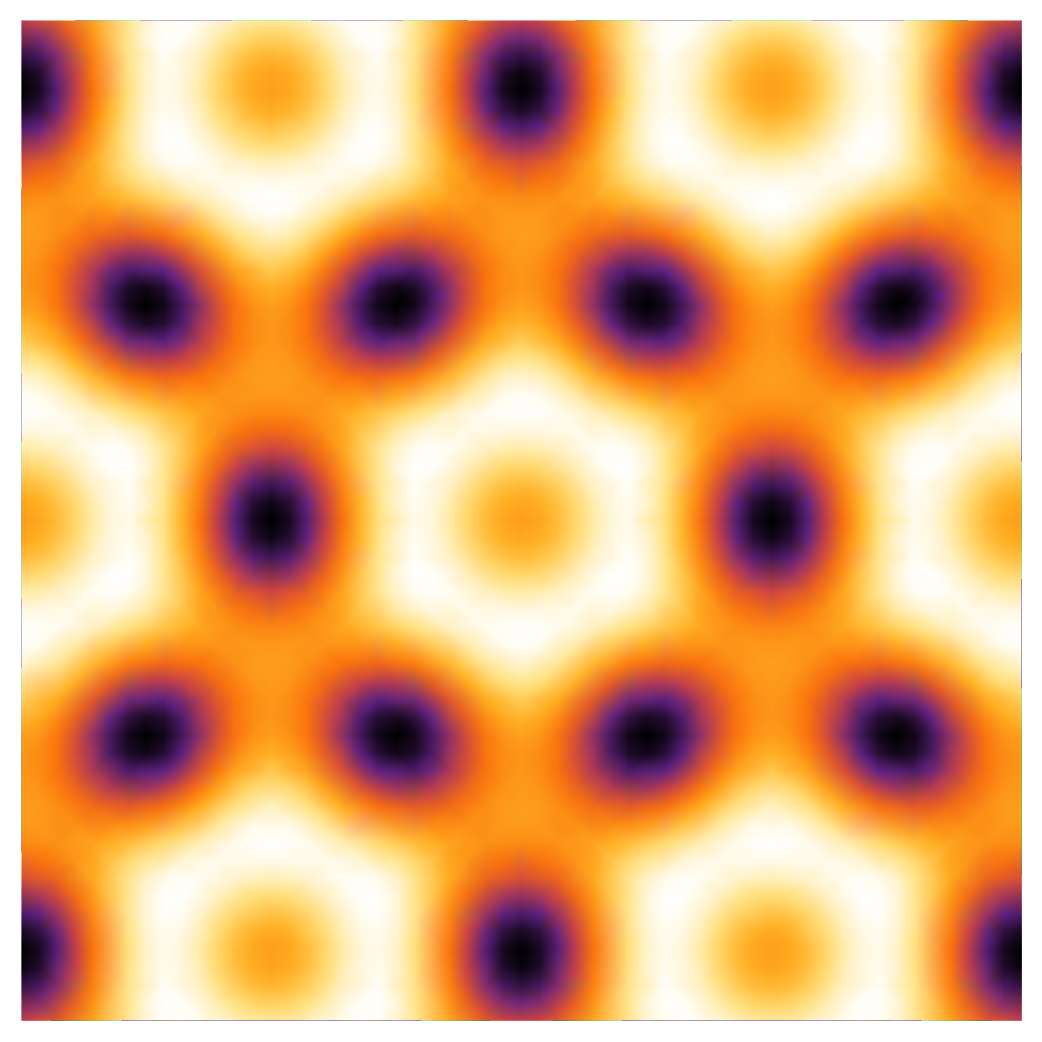}
    \begin{picture}(0,0)
    {\def\unitlength{}
        \put(-0.5\linewidth,1.15\linewidth){$(a)$}
    }
    \end{picture}
    \phantomsubcaption{\label{kagome}}
    \end{minipage}
    \begin{minipage}{0.15\textwidth}
    \includegraphics[width=\linewidth]{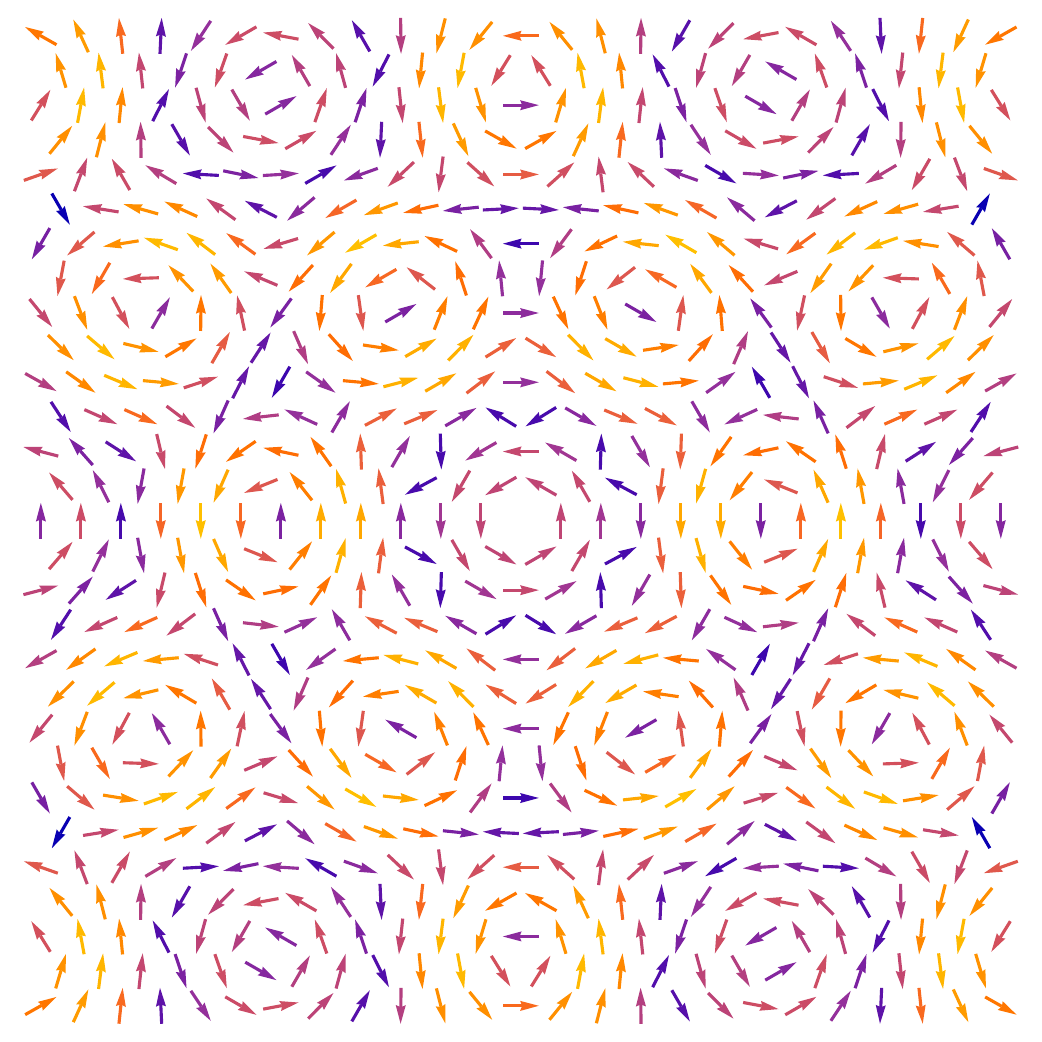}
    \begin{picture}(0,0)
    {\def\unitlength{}
        \put(-0.5\linewidth,1.15\linewidth){$(b)$}
    }
    \end{picture}
    \phantomsubcaption{\label{vector_potential}}
    \end{minipage}
    \begin{minipage}{0.15\textwidth}
    \includegraphics[width=\linewidth]{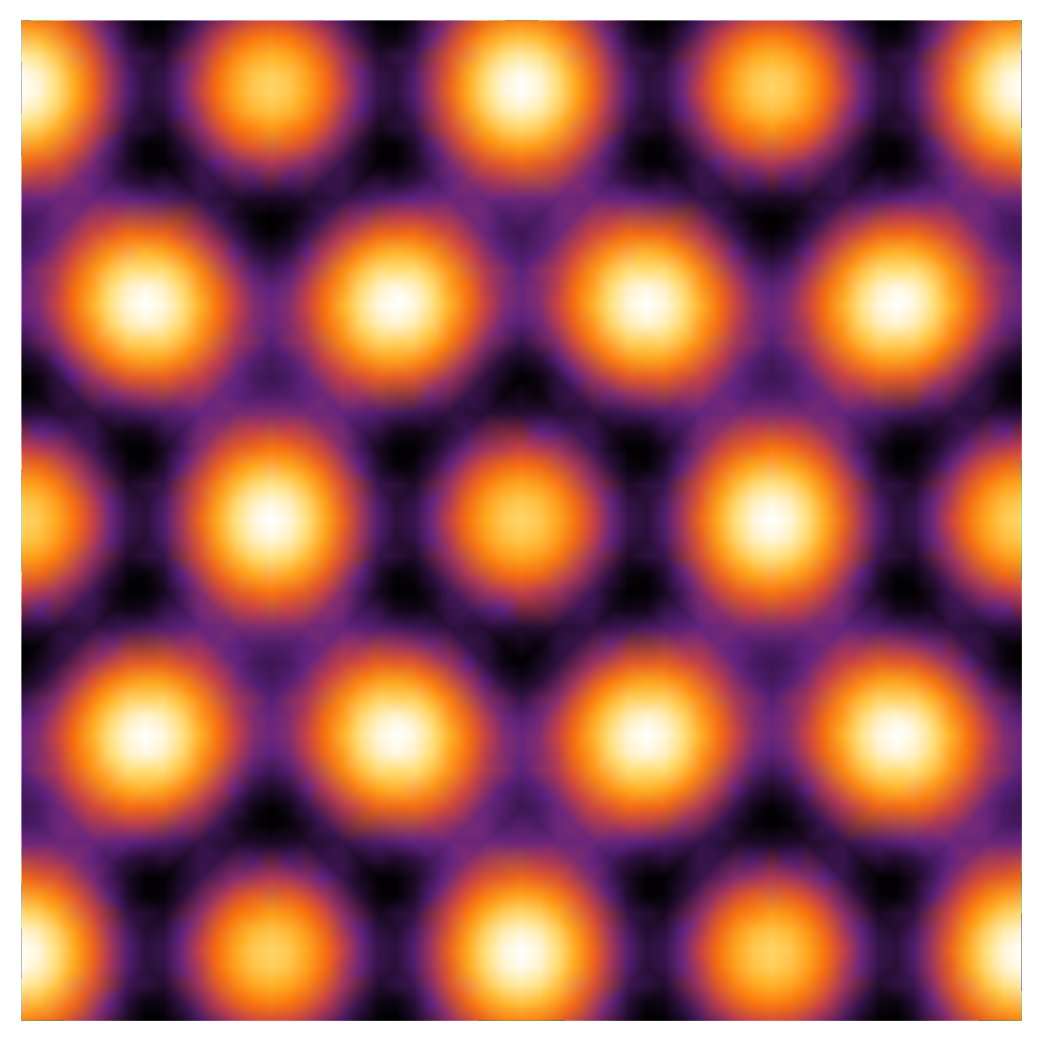}
    \begin{picture}(0,0)
    {\def\unitlength{}
        \put(-0.5\linewidth,1.15\linewidth){$(c)$}
    }
    \end{picture}
    \phantomsubcaption{\label{flux}}
    \end{minipage}
    \caption{
    \protect\subref{kagome}
    Kagome lattice potential from \cref{Vlattice}. The dark dots are the low-energy regions, which form the kagome lattice.
    \protect\subref{vector_potential}
    The effective gauge vector potential $\mathbf{A}_{\rm eff}$, the darker color being the stronger strength.
    \protect\subref{flux}
    The gauge field flux $\nabla \cross \mathbf{A}_{\rm eff}$, which is a scalar in 2d.
}
\end{figure}

We start by exploring a single-particle model describing electrons in a kagome lattice coupled to a single-mode gyrotropic cavity.
The coupling of the cavity vacuum to electrons generates an effective gauge field, which plays a crucial role in forming a CSL state. Notably, this interaction is induced entirely by the vacuum fluctuations of the cavity, without requiring external laser driving.
The full Hamiltonian is composed of three terms:
\begin{equation}
    H = H_{\rm k} + V_{\rm lattice}+H_{\rm cc}
\end{equation}
where $H_{\rm k}$ and $V_{\rm lattice}$ describe the kinetic energy and the kagome-lattice potential energy of electrons, and
$H_{cc}$ represents the gyrotropic cavity mode of energy $\omega_c$, \textit{i.e.}, $H_{\rm cc} = \omega_c a^{\dagger} a$ with photon annihilation and creation operator $a$ and $a^{\dagger}$.
The kinetic energy Hamiltonian, incorporating the cavity electromagnetic vector potential $\mathbf{A}$, is:
\begin{equation}\label{schrodingerpauli}
    H_{\rm k} = \frac{(\bm{\sigma} \cdot (\mathbf{p} - q\mathbf{A}))^2}{2m},
\end{equation}
where $\bm{\sigma}$ is the vector of Pauli matrices, $\mathbf{p}$ is the momentum of the electrons, $m$ is the mass of the electrons, the vector potential is $\mathbf{A} = A_0({\bm \epsilon} a + {\bm \epsilon}^* a^{\dagger})$
and, ${\bm \epsilon}=\frac{1}{\sqrt{2}}(1,i)$ represents the circularly polarized mode due to the gyrotropic cavity. Note that the Schr\"odinger-Pauli Hamiltonian \eqref{schrodingerpauli} automatically incorporates the interaction of the electron's spin with the magnetic field. The kagome lattice potential can be approximated by \cite{jo2012ultracold}:
\begin{equation} \label{Vlattice}
    V_{\rm lattice}(\mathbf{r}) = V_0 [\phi(\mathbf{r}) - \phi(2\mathbf{r})],
\end{equation}
where
\begin{equation}
    \phi(\mathbf{r}) = 
    \cos (\mathbf{b}_1 \cdot \mathbf{r})
    + \cos (\mathbf{b}_2 \cdot \mathbf{r})
    + \cos ((\mathbf{b}_1+\mathbf{b}_2) \cdot \mathbf{r})
\end{equation}
depends on the reciprocal vectors
$\mathbf{b}_1 = \frac{4\pi}{\sqrt{3}} (\frac{\sqrt{3}}{2},-\frac{1}{2})$,
$\mathbf{b}_2 = \frac{4\pi}{\sqrt{3}} (0,1)$.
Such a lattice potential is shown in \cref{kagome}.
As will be shown later, the microscopic details of the lattice potential do not alter our results qualitatively.

\paragraph{Asymptotically decoupled frame.}

To study the impact of cavity quantum fluctuations on the electronic system, we employ the asymptotically decoupled (AD) frame, a method originally used to study the cavity Lamb shift \cite{belov1989lamb} and more recently highlighted for its significance in the study of cavity materials \cite{ashida2021cavity}. The AD frame is obtained by performing a unitary transformation that effectively decouples the cavity photons from the electronic degrees of freedom, which becomes exact in either the weak ($A_0 \ll 1$) or strong ($A_0 \gg 1$) coupling limits. This transformation is given by:
\begin{equation}
   H_{\rm AD}= U^\dagger H U;~~{\text{with}}~U = \exp(-i \xi \mathbf{p} \cdot \bm\pi)
\end{equation}
where the parameter $\xi=\frac{qA_0}{m \tilde{\omega}_c}$ is of the length dimension,
$\tilde{\omega}_c = \omega_c + \frac{q^2 A_0^2}{m}$, $\bm{\pi} = i(\bm\epsilon^* a^{\dagger}-\bm\epsilon a)$ represents the dimensionless photon field momentum.

In the AD frame, the Hamiltonian splits into two parts: one representing the electrons in an effective gauge field (noting that this effective gauge field is distinct from the photon field) and the other representing the photon field itself. Ultimately, we can derive a second-quantized lattice model that effectively describes the system as detailed in the supplemental materials \cite{supp}
\begin{equation}\label{HAD}
\begin{aligned}
    H_{\rm AD} = &- \sum_{<i,j>,\sigma} te^{i\varphi_{ij}} c_{i\sigma}^{\dagger} c_{j\sigma} + U^' \sum_i n_{i\uparrow} n_{i\downarrow} \\
    &+ \sum_i \mathcal{B} (n_{i\uparrow} - n_{i\downarrow})
\end{aligned}
\end{equation}
where the phase factor $\varphi_{ij} = q\int_{\mathbf{r}_i}^{\mathbf{r}_j} \mathbf{A}_{\rm eff} \cdot d\mathbf{r}$ and $\mathbf{A}_{\rm eff} = \beta \hat{\mathbf{z}} \cross \nabla [\phi(\mathbf{r}) - \phi(2\mathbf{r})]$ is the effective gauge field coupled to the electrons, with dimensionless strength $\beta = \frac{m \xi^2 V_0}{2}$.
The onsite interaction $U^' \sum_i n_{i\uparrow} n_{i\downarrow}$ is also included, which can be shown to be invariant under the AD transformation.
In the last term, $\mathcal{B} = \frac{q^2A_0^2}{2m}$ acts as an effective Zeeman splitting field.

\begin{figure}[t]
    \centering
    \begin{minipage}{0.33\linewidth}
    \includegraphics[width=\linewidth, trim=1.9cm 1.9cm 1.5cm 1.5cm, clip]{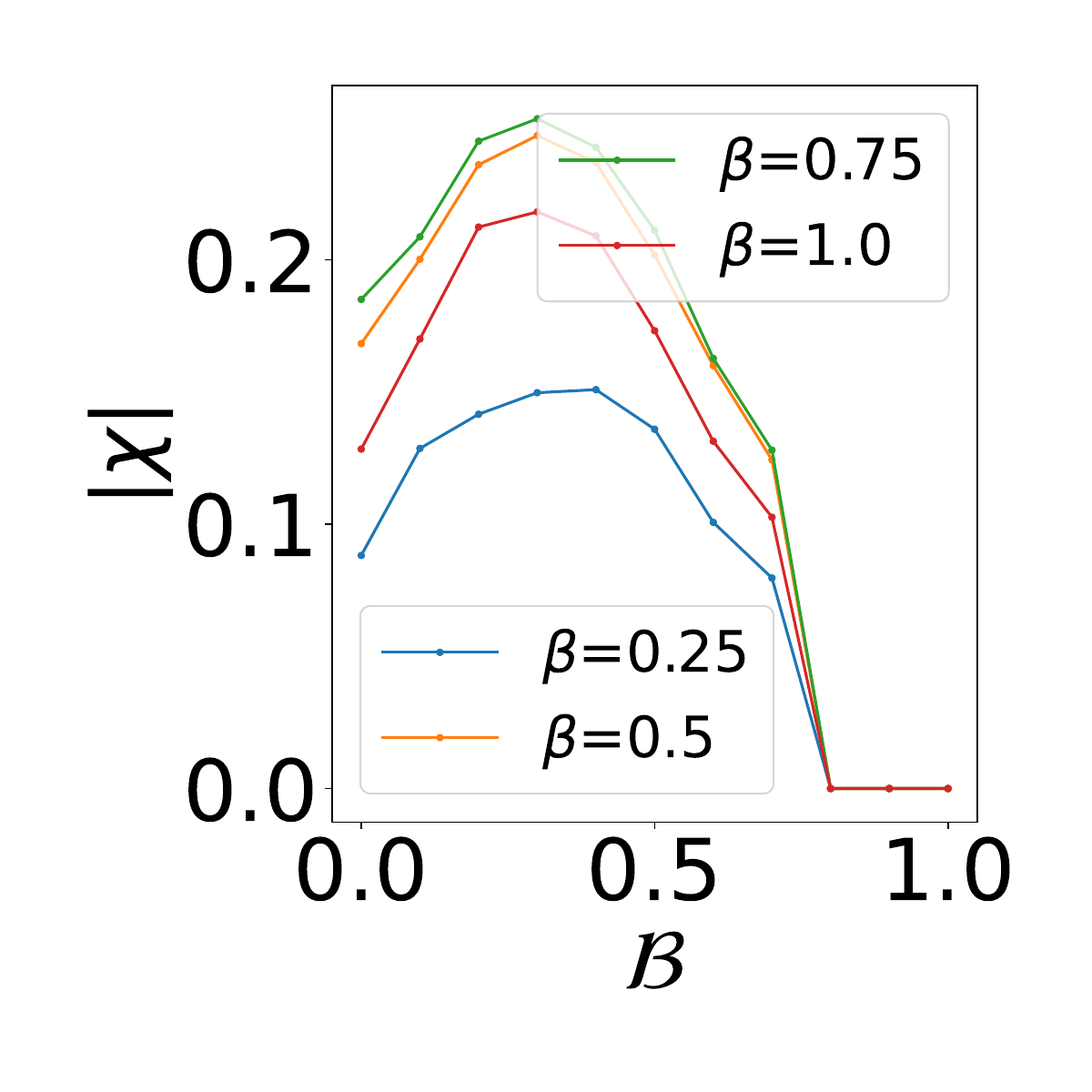}
    \begin{picture}(0,0)
    {\def\unitlength{}
        \put(-0.5\linewidth,1.05\linewidth){$(a)$}
    }
    \end{picture}
    \phantomsubcaption{\label{chiB}}
    \end{minipage}
    \begin{minipage}{0.32\linewidth}
    \includegraphics[width=\linewidth, trim=1.5cm 1.5cm 1.5cm 1.5cm, clip]{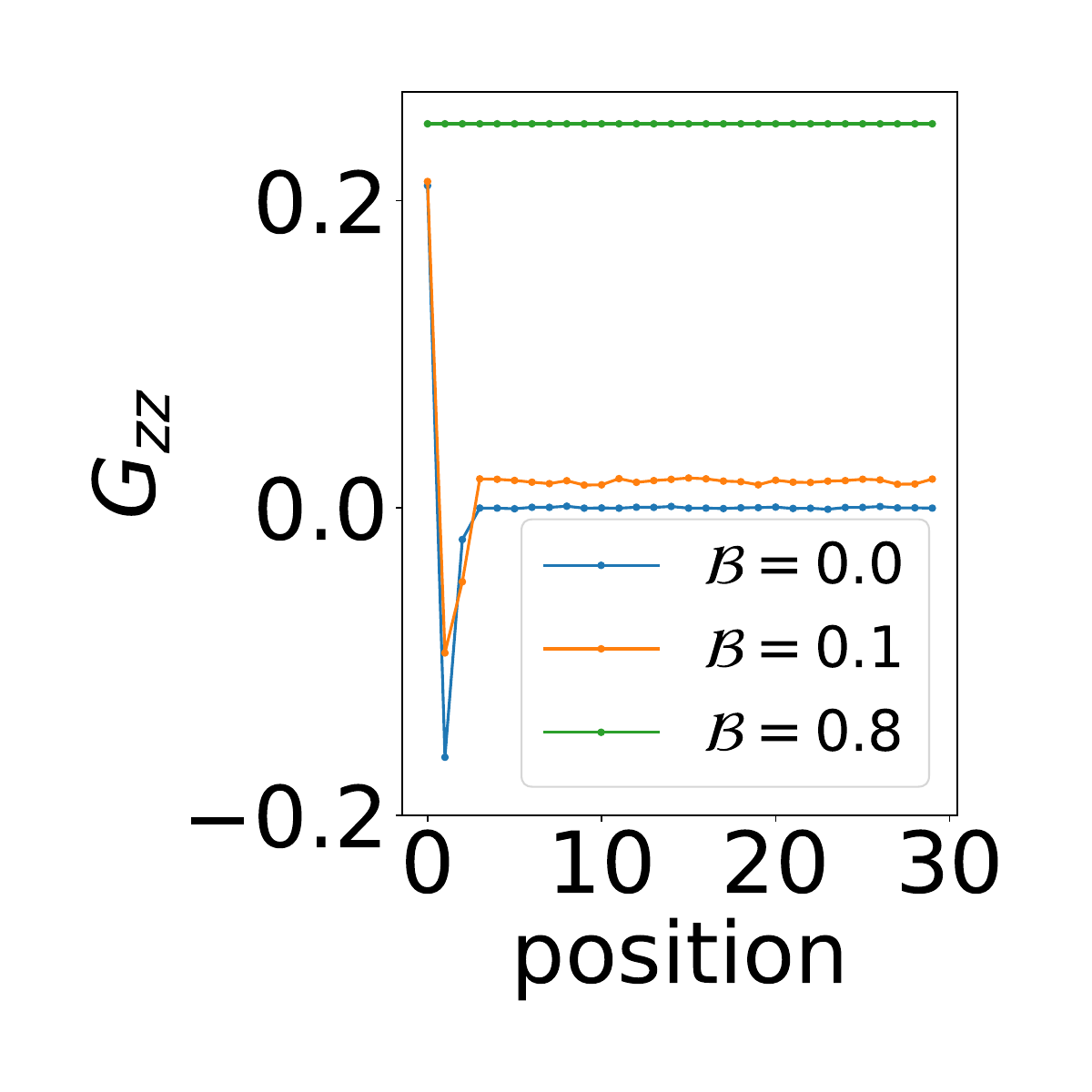}
    \begin{picture}(0,0)
    {\def\unitlength{}
        \put(-0.5\linewidth,1.0\linewidth){$(b)$}
    }
    \end{picture}
    \phantomsubcaption{\label{Gzz}}
    \end{minipage}
    \begin{minipage}{0.32\linewidth}
    \includegraphics[width=\linewidth, trim=1.5cm 1.5cm 1.5cm 1.5cm, clip]{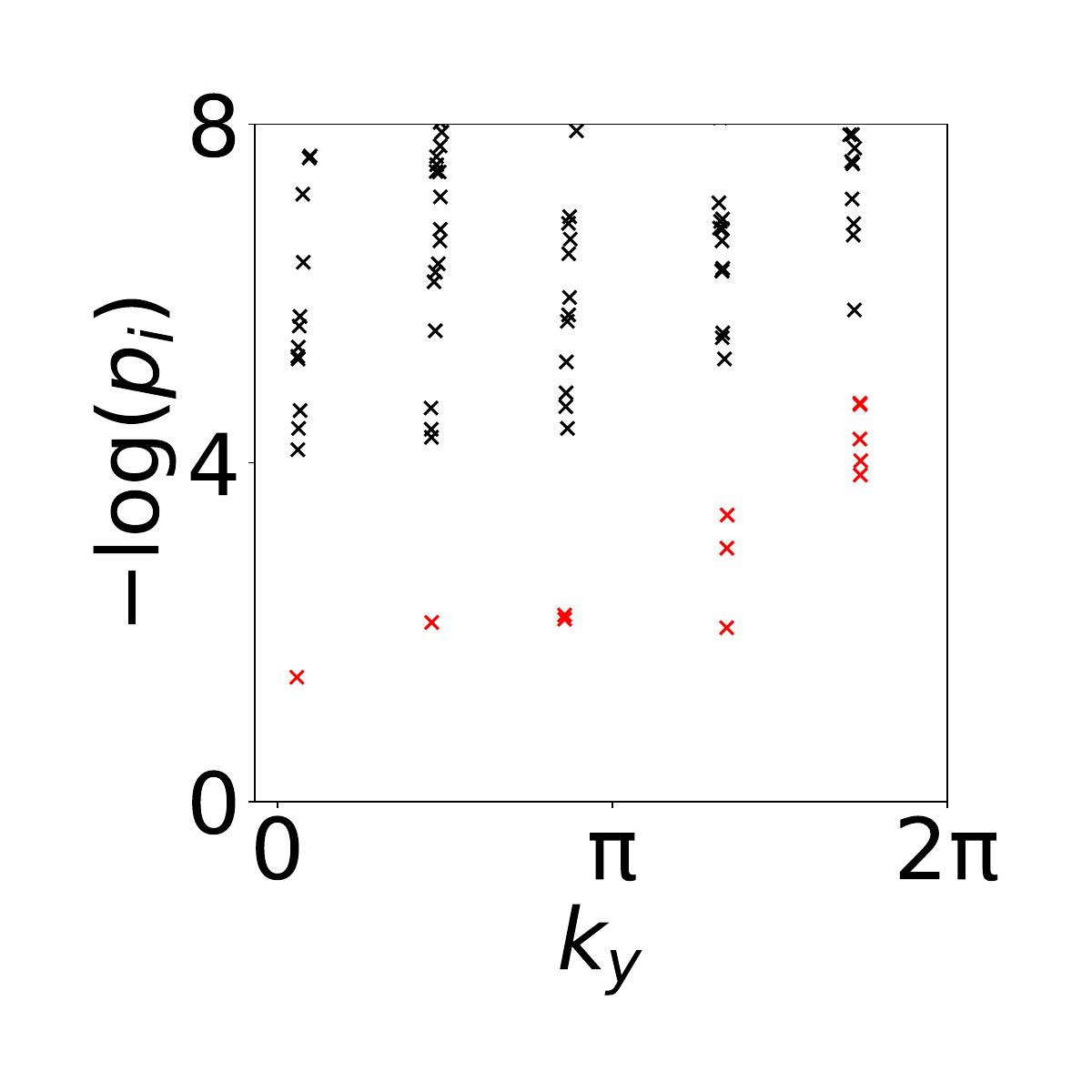}
    \begin{picture}(0,0)
    {\def\unitlength{}
        \put(-0.5\linewidth,1.0\linewidth){$(c)$}
    }
    \end{picture}
    \phantomsubcaption{\label{ES}}
    \end{minipage}
    \caption{\label{order}
    Order parameters for the kagome lattice inside the gyrotropic cavity.
    \protect\subref{chiB} is the chiral order parameter. DMRG simulation indicates a non-zero chiral order parameter for weak electron-photon interaction and a zero chiral order parameter for strong electron-photon interaction.
    $\beta = \frac{m \xi^2 V_0}{2}$
    is the dimensionless parameter describing the strength of the effective gauge field.
    \protect\subref{Gzz} is the spin $zz$ correlation when $\beta=0.5$. DMRG simulation indicates no or weak correlation for weak electron-photon interaction and a strong correlation for strong electron-photon interaction.
    Combining with the nonzero chiral order in the weak-coupling regime, this confirms the presence of a CSL phase, while strong coupling leads to a spin-polarized state.
    All the simulations above are done with $t=1$ and $U^'=5$.
    \protect\subref{ES} displays the entanglement spectrum with a cylinder geometry at small $\mathcal{B}$. The $k_y$ corresponds to the eigenvalue of translation in the wrapped direction $y$.
    The spectrum reveals a characteristic degeneracy pattern $1,1,2,3,5,\dots$ in the low-energy sector colored in red. This sequence reflects the underlying edge state described by a chiral bosonic CFT associated with the topological phase.
}
\end{figure}

By applying the AD unitary transformation and analyzing the resulting effective Hamiltonian, we observe how the coupling to a gyrotropic cavity vacuum state introduces a nontrivial phase factor $\varphi_{ij}$ in the electron hopping terms, the structure of which is shown in \cref{vector_potential,flux}. This phase factor is crucial, as it breaks time-reversal symmetry and ultimately leads to the realization of a chiral spin liquid. We confirm this behavior through simulations using single-site density matrix renormalization group (DMRG) of effective Hamiltonian \ref{HAD} with subspace expansion \cite{hubig2015strictly}. The simulations are performed with $10 \cross 10$ unit cells, energy truncation is $10^{-10}t$, and the bond dimension is 128 for CSL states. The results are shown in \cref{order}. In \cref{chiB}, the chiral order $\bar{\chi} = \langle \mathbf{S}_i \cdot \mathbf{S}_j \times \mathbf{S}_k \rangle$ remains non-zero when the electron-photon coupling $\mathcal{B}$ is smaller than a critical value. Beyond the critical value, the cavity quantum fluctuations polarize the spin and the chiral order vanishes, which is reflected in the spin $zz$ correlation $G_{zz}$ in \cref{Gzz}. By contrast, the $zz$ correlation is close to zero for smaller photon-electron coupling $\mathcal{B}$.
The absence of magnetic order and a non-zero chiral order confirms the CSL state of the system for weak photon-electron coupling.
Our simulations show that chiral order is preserved for weak photon-electron coupling, while strong coupling suppresses this order and polarizes the spin.

To further substantiate the topological nature of the state at small $\mathcal{B}$, we computed the entanglement spectrum, presented in \cref{ES}. Following the foundational analysis by Li and Haldane \cite{li2008entanglement}, the degeneracies observed in the low-energy sector of the entanglement spectrum are directly related to the underlying boundary conformal field theory (CFT). In particular, we observe a clear characteristic degeneracy sequence $1,1,2,3,5,\dots$, which precisely matches the counting of the chiral bosonic edge modes (see \cref{ES}). This sequence emerges due to the presence of gapless edge excitations that are governed by the chiral bosonic CFT intrinsic to this topological phase, thereby reinforcing our identification of the CSL state’s topological character.

\begin{table*}
\begin{tabular}{|c||c|c||c|c|}
\hline
& \multicolumn{2}{c||}{$\epsilon_r = 2$} & \multicolumn{2}{c|}{$\epsilon_r = 5$} \\
\hline
\diagbox{$\omega_c/2\pi \text{THz}$}{$a_0/\text{nm}$} & 0.1 & 0.5 & 0.1 & 0.5 \\
\hline \hline
0.1 & 
\begin{tabular}{@{}c@{}}$L/a_0 = 47.8$ \\ $V_0 = 0.328\,\text{eV}$\end{tabular} & 
\begin{tabular}{@{}c@{}}$L/a_0 = 28.0$ \\ $V_0 = 0.112\,\text{eV}$\end{tabular} & 
\begin{tabular}{@{}c@{}}$L/a_0 = 35.2$ \\ $V_0 = 0.178\,\text{eV}$\end{tabular} & 
\begin{tabular}{@{}c@{}}$L/a_0 = 20.6$ \\ $V_0 = 0.06\,\text{eV}$\end{tabular} \\
\hline
1 & 
\begin{tabular}{@{}c@{}}$L/a_0 = 22.2$ \\ $V_0 = 0.706\,\text{eV}$\end{tabular} & 
\begin{tabular}{@{}c@{}}$L/a_0 = 13.0$ \\ $V_0 = 0.241\,\text{eV}$\end{tabular} & 
\begin{tabular}{@{}c@{}}$L/a_0 = 16.4$ \\ $V_0 = 0.383\,\text{eV}$\end{tabular} & 
\begin{tabular}{@{}c@{}}$L/a_0 = 9.56$ \\ $V_0 = 0.131\,\text{eV}$\end{tabular} \\
\hline
10 & 
\begin{tabular}{@{}c@{}}$L/a_0 = 10.3$ \\ $V_0 = 1.52\,\text{eV}$\end{tabular} & 
\begin{tabular}{@{}c@{}}$L/a_0 = 6.03$ \\ $V_0 = 0.52\,\text{eV}$\end{tabular} & 
\begin{tabular}{@{}c@{}}$L/a_0 = 7.59$ \\ $V_0 = 0.826\,\text{eV}$\end{tabular} & 
\begin{tabular}{@{}c@{}}$L/a_0 = 4.44$ \\ $V_0 = 0.283\,\text{eV}$\end{tabular} \\
\hline
\end{tabular}
\caption{Cavity length $L_c$ when the transition happens and the lattice potential $V_0$ needed to observe the CSL phase is shown for a range of parameters, namely lattice constant $a_0$ and the relative permittivity $\epsilon_r$.}
\label{tab:experimental_parameters}
\end{table*}

A many-body version of AD transformation can also be performed and will result in a direct modification of the parameters as shown in the supplemental materials \cite{supp}
\begin{equation}
\begin{aligned}
    \tilde{\omega}_c &\to \widetilde{\omega}_c^{(MB)}=\omega_c(1+\frac{N q^2 A_0^2}{m \omega_c}), \\
\end{aligned}
\end{equation}
where $N$ is the number of total electrons.
For the purpose of experimental observation, we estimate that the transition occurs around $A_0 \approx \frac{\sqrt{mt}}{q}$.
Hence one gets the expression $A_0=\frac{\hbar}{q a_0}$, 
where the reduced Planck constant $\hbar$ and lattice constant $a$ are restored for clarity.
The amplitude $A_0$ is experimentally determined by cavity size, \textit{i.e.},
$
A_0 = \sqrt{\frac{\hbar}{2 \epsilon_r \epsilon_0 V \omega_c}}
$
where $\epsilon_0$ is the vacuum permittivity, $\epsilon_r$ is the relative permittivity, $V$ is the volume of the cavity, and $\omega_c$ is the cavity frequency that can be controlled experimentally.
One can then solve for the critical cavity length as shown in \cref{tab:experimental_parameters}.
For cubic cavities with dimensions exceeding $L_c$ -- the critical length scale, the system exhibits the CSL phase.
It's also useful to check the lattice potential depth when the CSL phase appears, namely $\beta \approx 0.1$, which gives rise to results in \cref{tab:experimental_parameters}.

\paragraph{Probe the photon and electron states with a waveguide.}

While previous proposals suggest that the CSL phase can be experimentally detected through edge currents, thermal Hall effect measurements, and other signatures of topological order, in the following, we show that, in a setup with a gyrotropic cavity, the photon states offer an additional pathway to probe the chiral spin states.
We return to the effective model \cref{HAD} in the AD frame to analyze experimental observables, focusing on the average photon number in the cavity and the associated transport properties when the system is coupled to a waveguide (see \cref{schematic}). These quantities provide insight into the emergent quantum state of the system induced by the cavity quantum fluctuations.
The averaged photon number $\bar{N} = \bra{g} a^{\dagger}a \ket{g}$, where $\ket{g}$ is the ground state of the system in the lab frame, can be evaluated in the AD frame as:
$
    \bar{N} 
    = \bra{0} \frac12 \xi^2 \mathbf{p}^2 \ket{0},
$
where $\ket{0}$ is the ground state in the AD frame,
indicating that the photon number is proportional to the averaged kinetic energy of the electronic state. This observation directly links the quantum state of the cavity photons to the electronic structure of the kagome lattice in a gyrotropic cavity.

\begin{figure}[t]
    \centering
    \begin{minipage}{0.49\linewidth}
    \includegraphics[width=\linewidth, trim=1.5cm 1.5cm 1.5cm 1.5cm, clip]{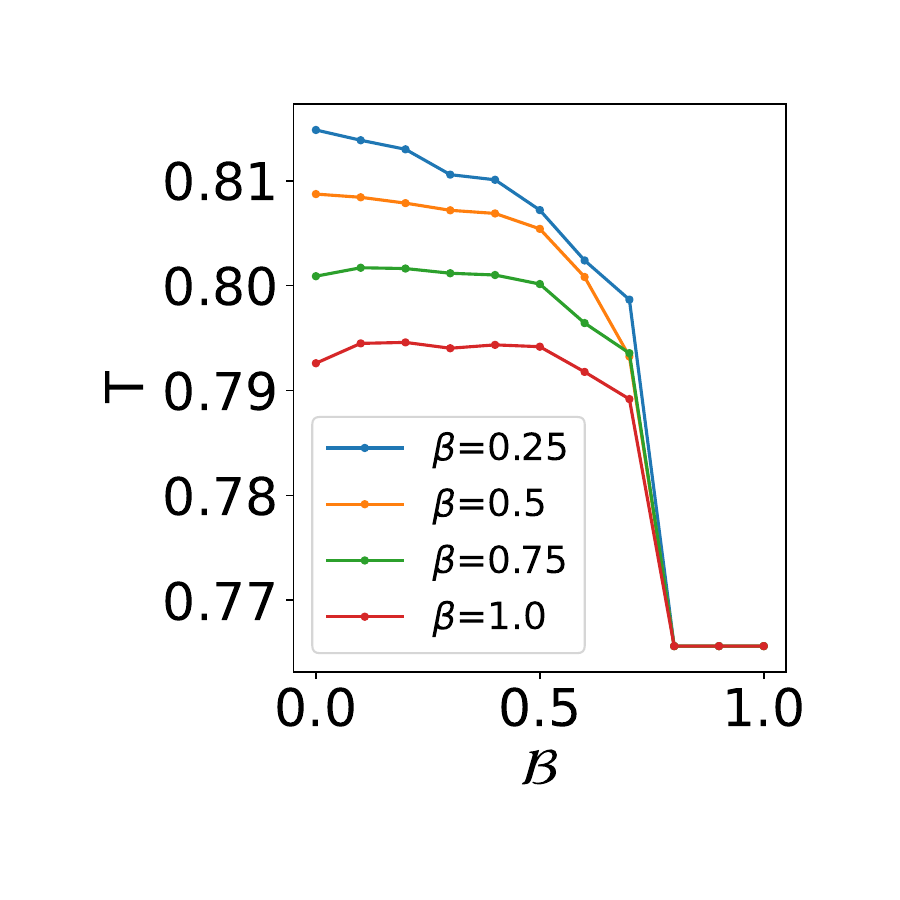}
    \begin{picture}(0,0)
    {\def\unitlength{}
        \put(-0.5\linewidth,1.0\linewidth){$(a)$}
    }
    \end{picture}
    \phantomsubcaption{\label{TB}}
    \end{minipage}
    \begin{minipage}{0.49\linewidth}
    \includegraphics[width=\linewidth, trim=1.5cm 1.5cm 1.5cm 1cm, clip]{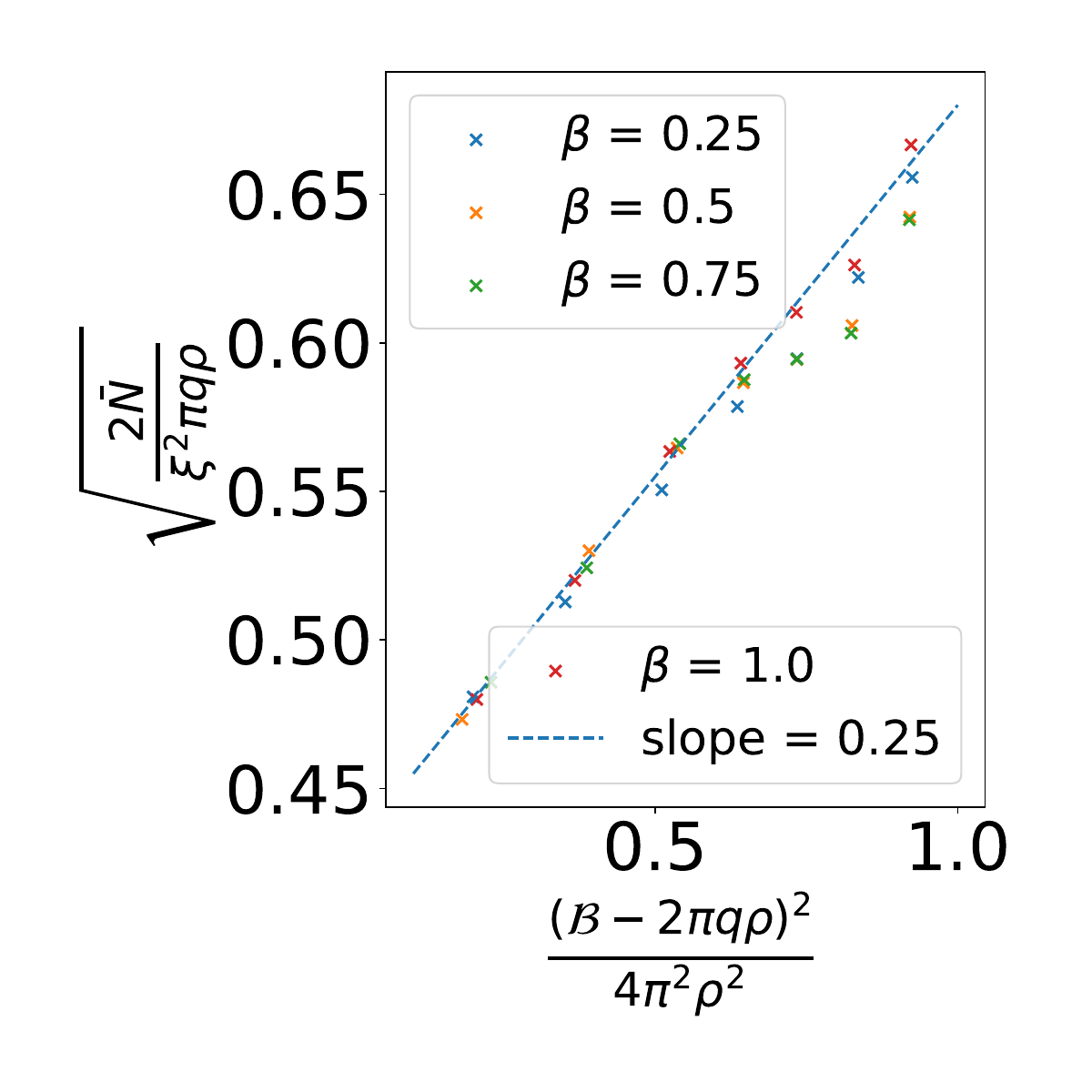}
    \begin{picture}(0,0)
    {\def\unitlength{}
        \put(-0.5\linewidth,1.0\linewidth){$(b)$}
    }
    \end{picture}
    \phantomsubcaption{\label{kappa}}
    \end{minipage}
    \caption{\label{transmission}
    \protect\subref{TB}
    shows the transmittance for a waveguide coupled to the {\color{blue}gyrotropic} cavity with a kagome lattice inside.
    The transmittance approaches a constant with strong electron-photon interaction. 
    All the simulations above are done with $t=1$, $U^'=5$, $\omega_k = 1$ and $g=0.1$.
    \protect\subref{kappa}
    shows the CS level fitting from the low-energy description of CSL. The results confirm the level-2 CS theory.
}
\end{figure}


Next, we examine the transport properties of photons through the waveguide coupled to the gyrotropic cavity.
The Hamiltonian governing photons in the cavity and in the waveguide is
\begin{equation}
    H_0 = \omega_c a^{\dagger} a +  \omega_k b_k^{\dagger} k_k
\end{equation}
where $\omega_c$ is the cavity frequency, and $\omega_k$ is the frequency of photons in an external waveguide of mode $k$. The coupling between the cavity photons and the waveguide photons may be described by
\begin{equation}
    H_g = g_k a^{\dagger} b_k + g_k^* b_k^{\dagger} a
\end{equation}
In the AD frame, the coupling Hamiltonian becomes:
\begin{equation}
    H_g^{(AD)} = g_k a^{\dagger} b_k + g_k^* b_k^{\dagger} a
    - g_k \xi \mathbf{p} \cdot \bm{\epsilon} b_k
    - g_k^* \xi \mathbf{p} \cdot \bm{\epsilon}^* b_k^{\dagger}.
\end{equation}
This modified Hamiltonian reflects how the photon-electron interaction affects the transport of photons between the cavity and the external waveguide. The additional terms proportional to $\xi \mathbf{p}$ represent the influence of the electrons’ momentum on the photon coupling. These terms induce a mixing between the cavity photons and the waveguide photons, mediated by the electron motion, and are responsible for modifying the transmittance of photons through the system.

The transmission amplitude can be calculated as shown in the supplemental materials \cite{supp}:
\begin{equation}
\begin{aligned}
    t_k &= \bra{g,k,t=+\infty}S\ket{g,k,t=-\infty} \\
    &= 1 - \frac12 \omega_k |g_k|^2
    - (\omega_k+\frac12) |g_k|^2 \bar{N}
\end{aligned}
\end{equation}
where $S = \mathcal{T} e^{-i\int dt (H_0 + H_g\delta(t))}$, $\xi=\frac{qA_0}{m \tilde{\omega}_c}$, $g_k$ is the coupling of the cavity and the waveguide, and $\omega_k$ is the dispersion of the waveguide.
The corresponding transmittance $T_k$ is given by $T_k = |t_k|^2$.
This equation shows how the transmission of photons through the system is reduced due to the interaction between the photons and the electrons in the kagome lattice. The reduction in transmission is proportional to the photon-electron coupling $|g_k|^2$ and the averaged photon number $\bar{N}$. This means that by measuring the transmittance, one can indirectly observe how the electron-photon interaction affects the electronic structure of the system. 
Particularly in the kagome lattice coupled to the gyrotropic cavity case, we show the DMRG results of the transmittance in \cref{TB}. When the electron-photon coupling is weak, namely the system is in a CSL state, the effective gauge field makes the transmittance $T_k$ a function of electron-photon coupling, while with strong electron-photon interaction, the polarized spin state is insensitive to the gauge field, leaving a constant transmittance.
In experimental setups, this observable provides a key signal of how the electronic state alters transport properties.

To further understand the connection between the transmittance behavior and the CSL states, we use a low-energy effective theory to describe the CSL state
\begin{equation}
    \begin{aligned}
        \mathcal{L} =& \psi^{\dagger} \left\{i\partial_{t}-qa_{0}-q\mathcal{A}_{0}
        + \frac{1}{2m} |\nabla - iq (\mathbf{a} + \pmb{\mathcal{A}})|^{2}\right\} \psi \\
        &+\frac{q^2\kappa}{4\pi} \epsilon^{\mu\nu p}a_{\mu}\partial_{\nu}a_{\rho}
        +\frac{q^2}{2\pi}\epsilon^{\mu\nu p}\mathcal{A}_{\mu}\partial_{\nu}a_{\rho},
    \end{aligned}
\end{equation}
where $\psi$ is a bosonic mode, $a_{\mu}$ is an emergent Chern-Simons (CS) gauge field with level-$\kappa$ and $\mathcal{A}_{\mu}$ is the external field.
As detailed in \onlinecite{supp}, the averaged photon number can be written as
\begin{equation} \label{Nbar-kappa}
    \bar{N}
    =\frac{1}{2}\xi^{2} \pi q\rho \left[ 1 + \frac{(\mathcal{B}-2\pi q\rho)^2}{4\pi^2 \kappa^2 \rho^2} \right]^2,
\end{equation}
with $\rho = \langle \psi^{\dagger} \psi \rangle$ can be estimated by magnetization $\langle S_z \rangle + 1/2$.

To confirm this behavior, we compute the quantity $\sqrt{\frac{2\bar{N}}{\xi^2\pi q\rho}}$ using the DMRG ground state and plot it against $\frac{(\mathcal{B}-2\pi q\rho)^2}{4\pi^2 \rho^2}$. According to \cref{Nbar-kappa}, these two quantities should exhibit a linear relationship with a slope equal to $\frac{1}{\kappa^2}$. For the CSL on the kagome lattice, this slope equals $0.25$, consistent with the numerical results shown in \cref{kappa} at small values of $\mathcal{B}$. However, as $\mathcal{B}$ increases, the numerical results deviate from the low-energy theoretical prediction.

\paragraph{Discussion and Summary.}

This study demonstrates how coupling a kagome lattice to a gyrotropic cavity field can realize a CSL phase, providing both a theoretical foundation and practical insights for tuning quantum materials using cQED. By leveraging the interplay between the frustrated geometry of the kagome lattice and cavity-induced chiral interactions, we show how the effective spin model at half-filling leads to the emergence of CSLs.
We also discuss the experimental feasibility, highlighting recent cQED advances that enable the creation of tunable gyrotropic cavities. Platforms such as kagome-based quantum magnets or ultracold atomic lattices in optical cavities offer promising avenues for realizing the predicted CSL phase \cite{jo2012ultracold}.
In the cold atom setup, the Feshbach resonance can be employed to tune the interaction, and by controlling the atom number and adiabatically loading them into the optical lattice \cite{jordens2008mott}.
Herbertsmithite is another example of a material that realizes the kagome lattice.
Its half-filled electronic structure, combined with significant onsite interactions, has been extensively studied as a potential candidate for exhibiting QSL state\cite{norman2016herbertsmithite}.
Experimental observables, including photon number and transport, that are quantized according to the level of emergent CS theory,
provide clear signatures for detecting and manipulating CSL states in strongly correlated systems, opening new possibilities for their exploration and control.

Looking forward, this work opens several intriguing research directions. One of the most exciting possibilities is the study of doped CSLs, where the system is not restricted to half-filling. In such cases, the introduction of charge carriers could couple the charge degrees of freedom to the CSL state, potentially leading to new and exotic quantum phases. Doping a CSL can also break spin-charge separation, resulting in rich phase diagrams that may include superconducting phases or charge-density wave states. Studies in this context, such as those investigating fractionalized Fermi liquids and other strongly correlated doped CSLs, suggest that novel topological states could emerge from the interplay between charge fluctuations and spin chirality \cite{grover2010weak,savary2017superconductivity}.

In conclusion, the results presented here establish a solid framework for designing and probing CSLs within a practical gyrotropic cavity setup, paving the way for further exploration of doped CSLs, cavity-tuned phase transitions, and nonequilibrium quantum phases in strongly correlated systems. Building on recent experimental advances in cavity-based setups \cite{chiocchetta2021cavity,schlawin2022cavity,ashida2021cavity}, the intrinsic flexibility of cQED allows for precise control over interaction strengths and cavity parameters, enabling the engineering of phase transitions between various topological quantum phases. This approach also opens the possibility of harnessing cavity quantum fluctuations to create exotic excitations with fractional statistics, a crucial step toward advancing topological quantum computing.

\begin{acknowledgements}
{\it Acknowledgements---}This work was supported
by the Innovation Program for Quantum Science and
Technology Grant No. 2021ZD0301900, National Natural Science Foundation of China (NSFC) under Grant No.
12374332, Project supported by Cultivation Project of
Shanghai Research Center for Quantum Sciences Grant
No. LZPY2024, Shanghai Science and Technology Innovation Action Plan Grant No. 24LZ1400800. C.W. was partially supported by Higher Education and Science Committee of MESCS RA (Research Project No. 25PostDoc1C003).
\end{acknowledgements}


\section{End Matter}

\paragraph{From cavity photon-electron coupling to chiral interaction.}

To understand how the coupling to gyrotropic cavity photons induces a CSL phase, it is pedagogical to derive the effective Hamiltonian of the system in terms of spins, particularly in the strong-coupling regime where the kinetic energy of electrons is much smaller compared to the interaction energy. This regime can be analyzed through the $t/U$ expansion \cite{takahashi1977half,macDonald1988expansion,motrunich2006orbital} of the Hubbard model \cref{HAD} at half-filling.
This technique allows us to derive an effective spin Hamiltonian that captures the low-energy physics of the system. Notably, the gyrotropic cavity introduces nontrivial phase factors in the hopping terms of the effective Hamiltonian, leading to chiral three-spin interactions. These interactions are essential for breaking time-reversal symmetry and stabilizing the CSL phase.
In this limit, the effective Hamiltonian becomes:
\begin{equation}
    \begin{aligned}
        H_{\rm eff} =
        & \sum_{\text{\tikz[baseline=-0.5ex]
        {
        \coordinate (A) at (0,0);
        \coordinate (B) at (0.5,0);
        \fill (A) circle (2pt);
        \fill (B) circle (2pt);
        \draw (A) -- (B);
        }}}
        \frac{2|t_{ij}|^2}{U^'} (2 \mathbf{S}_i \cdot \mathbf{S}_j - \frac12) \\
        & -
        \sum_{\text{\tikz[baseline=-0.5ex]
        {
        \coordinate (A) at (0,0);
        \coordinate (B) at (0.25,0.433);
        \coordinate (C) at (0.5,0);
        \fill (A) circle (2pt);
        \fill (B) circle (2pt);
        \fill (C) circle (2pt);
        \draw (A) -- (B) -- (C) -- (A);
        }}}
        \frac{24\Im(t_{ij} t_{jk} t_{ki})}{U^{'2}} \mathbf{S}_i \cdot (\mathbf{S}_j \times \mathbf{S}_k)
    \end{aligned}
\end{equation}
where the site-dependent hoppings are $t_{ij} = te^{i\varphi_{ij}}$,
the first term describes Heisenberg-like interactions between neighboring spins,
and the second term introduces a chiral three-spin interaction, which is summed over all the triangles in the kagome lattice.
The chiral nature of these terms stems from the nontrivial phase factors induced by the cavity mode, leading to the spin chirality operator,
which promotes a net spin current around triangular plaquettes of the kagome lattice.

This effective spin model reveals the crucial role of the phase factor induced by the gyrotropic cavity field in generating chiral spin correlations. Both theoretical work \cite{bauer2014chiral,bauer2014gapped,kumar2015chiral,maiti2019fermionization,niu2022chiral} and experimental evidence \cite{claassen2017dynamical,schweika2022chiral} have shown that such chiral interactions are essential for stabilizing a CSL phase.

\paragraph{Phase diagram for experimental exploration}

\begin{figure}[t]
    \centering
    \begin{minipage}{\linewidth}
    \includegraphics[width=\linewidth]{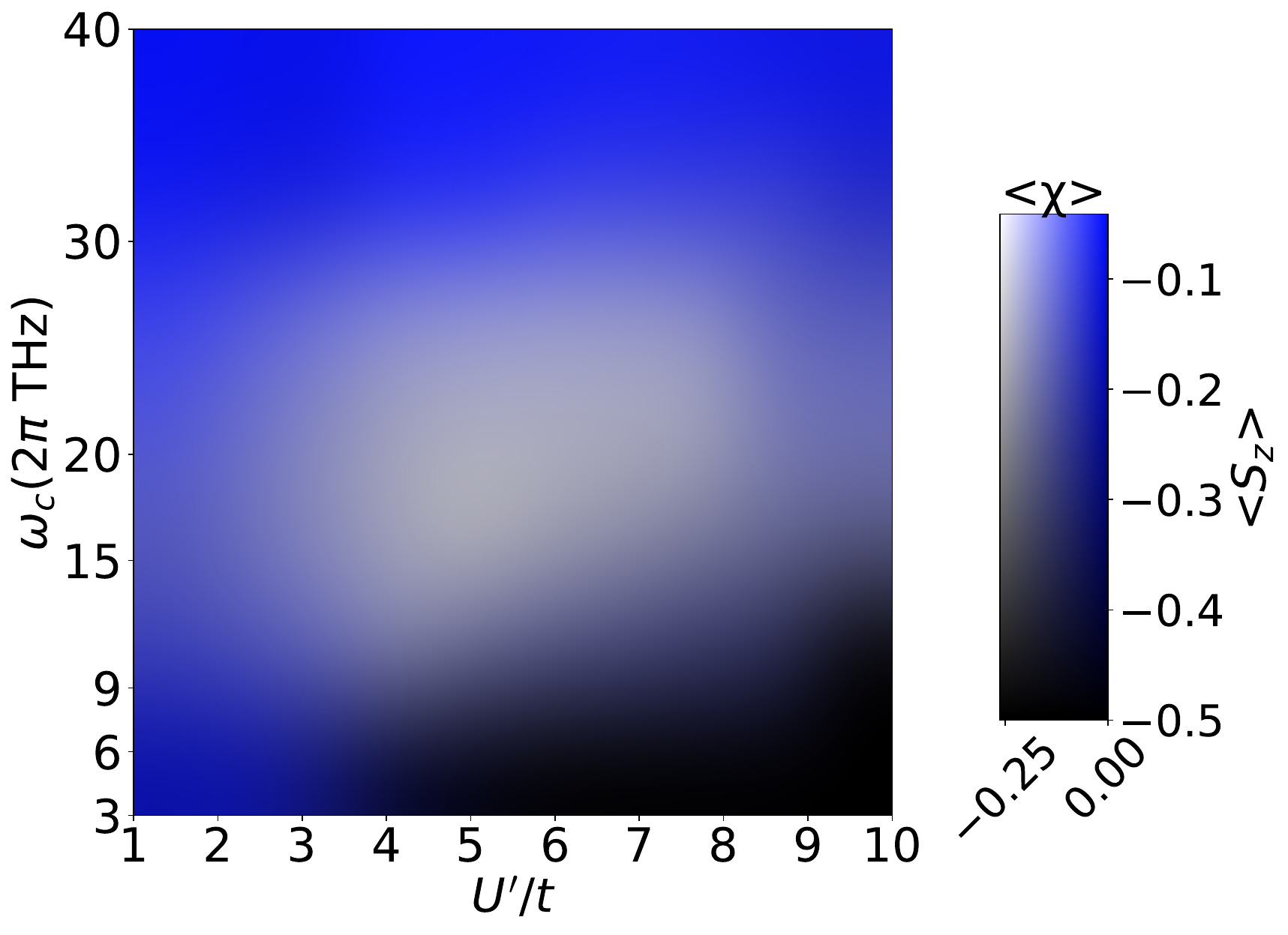}
    \begin{picture}(0,0)
    {\def\unitlength{}
        \put(-0.12\linewidth,0.4\linewidth){\color{yellow}CSL}
        \put(-0.05\linewidth,0.16\linewidth){\color{yellow}polarized state}
        \put(-0.30\linewidth,0.65\linewidth){\color{yellow}metallic phase or QSL }
        \put(-0.35\linewidth,0.19\linewidth){\color{yellow}metallic}
        \put(-0.35\linewidth,0.16\linewidth){\color{yellow}phase}
    }
    \end{picture}
    \end{minipage}
    \caption{\label{phase}A phase diagram computed using typical herbertsmithite experimental parameters. The saturation and brightness of the color represent the average magnetization and chirality separately. For small $U'$, the blue region is a band metallic phase with vanishing chirality and total magnetization. With small cavity frequency and large $U'$, the effective Zeeman term dominates and the state is polarized. As one increases the cavity frequency in the large $U'$ region, the system enters into the CSL phase with finite chirality. For super large cavity frequency, the photonic modes are effectively gapped, and the system becomes a clean kagome lattice, which potentially supports other QSLs.
}
\end{figure}

To facilitate further experimental investigation, we calculated the phase diagram using parameters for the material herbertsmithite: a nearest-neighbor hopping strength of $t_1 = 17 \rm~meV$, a next-nearest-neighbor hopping of $t_2 = 0.29 \rm~meV$, and a lattice constant of $a = 6.834 \text{ \AA}$.
The hopping $t_1$ sets the mass term $m$ in our model, which should not be confused with the hopping $t$ in the effective Hamiltonian.
The resulting phase diagram, presented in \cref{phase}, reveals a CSL phase, characterized by finite chirality and small magnetization (shaded grey). Our calculations indicate that this CSL phase is stabilized at cavity frequencies around 20 THz — a regime promisingly accessible in LC circuit cavities employed in recent experiments \cite{appugliese2022breakdown,enkner2025tunable}.

%

\onecolumngrid 
\clearpage     



\section*{Supplementary material}

# Supplemental Material for Cavity-Vacuum-Induced Chiral Spin Liquids in Kagome Lattices: Tuning and Probing Topological Quantum Phases via Cavity Quantum Electrodynamics

Chenan Wei,[1,2] Liu Yang,[3,4] and Qing-Dong Jiang[3,4,5]

[1]*Department of Physics, University of Massachusetts, Amherst, Massachusetts 01003, USA*

[2]*Department of Physics, University of Alberta, Edmonton, Alberta T6G 2E1, Canada*

[3]*Tsung-Dao Lee Institute, Shanghai Jiao Tong University, Shanghai 200240, China*

[4]*School of Physics and Astronomy, Shanghai Jiao Tong University, Shanghai 200240, China*

[5]*Shanghai Branch, Hefei National Laboratory, Shanghai 201315, China*

(Dated: November 12, 2024)

In this supplemental material, we will present the detailed derivation of the equations used in the main text.

## I. KAGOME LATTICE COUPLED TO A CHIRAL CAVITY IN THE ASYMPTOTIC DECOUPLED FRAME

### A. A single particle version

We start with the Hamiltonian describing electrons in a kagome lattice coupled to a single-mode chiral cavity presented in the main text

$$H = \frac{(\boldsymbol{\sigma} \cdot (\mathbf{p} - q\mathbf{A}))^2}{2m} + V_{\text{lattice}}(\mathbf{r}) + \omega_c a^\dagger a, \tag{1}$$

where $\boldsymbol{\sigma}$ is the vector of Pauli matrices, $\mathbf{p}$ is the momentum of the electrons, $m$ is the mass of the electrons, the vector potential is $\mathbf{A} = A_0(\boldsymbol{\epsilon} a + \boldsymbol{\epsilon}^* a^\dagger)$ and $\boldsymbol{\epsilon} = \frac{1}{\sqrt{2}}(1, i)$, which is circularly polarized because of the chiral cavity, $a$ and $a^\dagger$ are photon annihilation and creation operators.

To simplify the Hamiltonian 1, the following two relations can be used

$$(\boldsymbol{\sigma} \cdot \mathbf{a})(\boldsymbol{\sigma} \cdot \mathbf{b}) = \mathbf{a} \cdot \mathbf{b} + i\boldsymbol{\sigma} \cdot (\mathbf{a} \times \mathbf{b}), \tag{2}$$

and

$$[A_x, A_y] = -iA_0^2. \tag{3}$$

With the above relations, one can expand the Hamiltonian 1 and gets

$$H = \frac{\mathbf{p}^2}{2m} - \frac{q\mathbf{p} \cdot \mathbf{A}}{m} - \frac{q\boldsymbol{\sigma} \cdot \mathbf{B}}{2m} + V_{\text{lattice}}(\mathbf{r}) + \tilde{\omega}_c a^\dagger a + \frac{q^2 A_0^2 \sigma_z}{2m}, \tag{4}$$

where $\tilde{\omega}_c = \omega_c + \frac{q^2 A_0^2}{m}$, and $B = \partial_x A_y - \partial_y A_x$. If we consider the system at the peak of the vector potential $\mathbf{A}$, the magnetic field vanishes and hence we can remove the term $-\frac{q\boldsymbol{\sigma} \cdot \mathbf{B}}{2m}$.

To decouple the cavity photons from the electronic degrees of freedom, the asymptotically decoupled(AD) transformation is employed, which is a unitary transformation

$$U = \exp(-i\xi \mathbf{p} \cdot \boldsymbol{\pi}) \tag{5}$$

where the parameter $\xi = \frac{qA_0}{m\tilde{\omega}_c}$ is chosen such that the coupling between the electron and photon vanishes, and $\boldsymbol{\pi} = i(\boldsymbol{\epsilon}^* a^\dagger - \boldsymbol{\epsilon} a)$ represents the photon field momentum.

To calculate the transformed Hamiltonian

$$H_U = U^\dagger H U = \exp(i\xi \mathbf{p} \cdot \boldsymbol{\pi}) H \exp(-i\xi \mathbf{p} \cdot \boldsymbol{\pi}), \tag{6}$$

the Hadamard formula is used

$$e^{\hat{A}}\hat{B}e^{-\hat{A}} = \hat{B} + [\hat{A}, \hat{B}] + \frac{1}{2!}[\hat{A}, [\hat{A}, \hat{B}]] + \frac{1}{3!}[\hat{A}, [\hat{A}, [\hat{A}, \hat{B}]]] + \dots \tag{7}$$

One can verify that, under the AD transformation

$$U^\dagger \mathbf{A} U = A_0 \left[ \boldsymbol{\epsilon}^* a^\dagger + \boldsymbol{\epsilon} a + \xi(\mathbf{p} \cdot \boldsymbol{\epsilon}^*)\boldsymbol{\epsilon} + \xi(\mathbf{p} \cdot \boldsymbol{\epsilon})\boldsymbol{\epsilon}^* \right], \tag{8}$$

$$U^\dagger a^\dagger a U = a^\dagger a + \xi \mathbf{p} \cdot (\boldsymbol{\epsilon}^* a^\dagger + \boldsymbol{\epsilon} a) + \xi^2 (\mathbf{p} \cdot \boldsymbol{\epsilon})(\mathbf{p} \cdot \boldsymbol{\epsilon}^*) \tag{9}$$

$$U^\dagger \mathbf{r} U = \mathbf{r} + \xi \boldsymbol{\pi} - \frac{1}{2}\xi^2 \mathbf{p} \times \hat{\mathbf{z}}. \tag{10}$$

With the above equations, the transformed Hamiltonian, under the photon vacuum subspace, turns out to be

$$\langle 0| H_U |0\rangle \approx \frac{\mathbf{p}^2}{2\tilde{m}} + V_{\text{lattice}}(\mathbf{r} - \frac{1}{2}\xi^2 \mathbf{p} \times \hat{\mathbf{z}}) + \frac{q^2 A_0^2 \sigma_z}{2m} \tag{11}$$

where $\frac{1}{\tilde{m}} = \frac{1}{m}(1 - \frac{q^2 A_0^2}{m\tilde{\omega}_c})$, namely $\tilde{m} = m + \frac{q^2 A_0^2}{\omega_c}$.

The lattice potential can be further simplified

$$V_{\text{lattice}}(\mathbf{r} - \frac{1}{2}\xi^2 \mathbf{p} \times \hat{\mathbf{z}}) \approx V_{\text{lattice}}(\mathbf{r}) - \frac{1}{2}\xi^2 \mathbf{p} \cdot (\hat{\mathbf{z}} \times \nabla V_{\text{lattice}}(\mathbf{r})). \tag{12}$$

The second term works as an effective gauge field

$$\mathbf{A}_{\text{eff}} = \frac{\tilde{m}\xi^2}{2}\hat{\mathbf{z}} \times \nabla V_{\text{lattice}}(\mathbf{r}). \tag{13}$$

As a result, the Hamiltonian in AD frame becomes

$$H_U = \frac{(\mathbf{p} - q\mathbf{A}_{\text{eff}})^2}{2\tilde{m}} + V_{\text{eff}}(\mathbf{r}) + \frac{q^2 A_0^2 \sigma_z}{2m}, \tag{14}$$

where the effective potential

$$V_{\text{eff}}(\mathbf{r}) = V_{\text{lattice}}(\mathbf{r}) - \frac{\tilde{m}\xi^4}{8}(\nabla V_{\text{lattice}}(\mathbf{r}))^2. \tag{15}$$

With these preparation, one can straightforwardly write down the lattice model in the AD frame

$$H_{\text{AD}} = -\sum_{<i,j>,\sigma} t e^{i\varphi_{ij}} c_{i\sigma}^\dagger c_{j\sigma} + U' \sum_i n_{i\uparrow} n_{i\downarrow} + \sum_i \frac{q^2 A_0^2 \sigma_z}{2m}(n_{i\uparrow} - n_{i\downarrow}) \tag{16}$$

as presented in the main text. The phase factor can be computed from the effective gauge field $\varphi_{ij} = q \int_{\mathbf{r}_i}^{\mathbf{r}_j} \mathbf{A}_{\text{eff}} \cdot d\mathbf{r}$.

### B. A many-body version

Now, if we consider the mediated interaction between electrons, we can derive the AD Hamiltonian from a many-body perspective. We now write down the Hamiltonian for all the electrons as follows:

$$\mathcal{H} = \sum_{j=1}^N \left( \frac{\mathbf{p}_j^2}{2m} - \frac{q\mathbf{p}_j \cdot \mathbf{A}}{m} + V_{\text{lattice}}(\mathbf{r}_j) + \frac{q^2 A_0^2 \sigma_z^j}{2m} \right) + \omega_c(a^\dagger a + \frac{1}{2}), \tag{17}$$

where we have removed the interaction $\sigma \cdot \mathbf{B}$ by considering the material layer at the peak of the vector potential. Now we apply the unitary transformation

$$\mathcal{U} = \exp\left[ -i\xi(\sum_j \mathbf{p}_j) \cdot \hat{\boldsymbol{\pi}} \right], \tag{18}$$

where

$$\xi = \frac{qA_0}{m\widetilde{\omega}_c^{(MB)}}, \tag{19}$$

with

$$\widetilde{\omega}_c^{(MB)} = \omega_c(1 + \frac{Nq^2 A_0^2}{m\omega_c}), \tag{20}$$

to the many-body Hamiltonian and obtain

$$\begin{aligned}
\mathcal{H}_U &= \mathcal{U}^\dagger \mathcal{H} \mathcal{U} \\
&= \left[ \sum_{j=1}^N \frac{\mathbf{p}_j^2}{2m} + V_{\text{lattice}}(\mathbf{r}_j + \boldsymbol{\tau}_j) + \frac{q^2 A_0^2 \sigma_z^j}{2m} \right] - \frac{\widetilde{\omega}_c^{(MB)}}{2}(\sum_{j=1}^N \xi \mathbf{p}_j)^2 + \widetilde{\omega}_c^{(MB)}(a^\dagger a + \frac{1}{2}).
\end{aligned} \tag{21}$$

Here, we define a new operator $\boldsymbol{\tau}_j$ as

$$\boldsymbol{\tau}_j = \xi\hat{\boldsymbol{\pi}} - \frac{1}{2}\xi^2 \mathbf{p}_j \times \hat{\mathbf{z}}. \tag{22}$$

The term

$$\mathcal{H}' = -\frac{\widetilde{\omega}_c^{(MB)}}{2}(\sum_{j=1}^{N} \xi \mathbf{p}_j)^2 \tag{23}$$

contains an all-to-all interaction. In fact, the all-to-all cavity-mediated interaction can be reduced to an electron-independent Hamiltonian because the system is uniform on a large scale [1]. Specifically, every particle experiences an identical average effect from the rest, effectively averaging out the microscopic interactions. To perform the electron-independent approximation, we first take the expectation of $\mathcal{H}'$ under a many-body wave function as follow $\Psi(x_1,..,x_N)$:

$$\mathcal{H}'[\Psi] = -\frac{\widetilde{\omega}_c^{(MB)}\xi^2}{2}\int d\Gamma \, \Psi^*(x)(\sum_{j=1}^{N} \mathbf{p}_j)^2\Psi(x), \tag{24}$$

Here the integrated variable $d\Gamma = \Pi_i dx_i$. Now, we introduce the single-body and two-body wave functions $\Psi_1(x)$ and $\Psi_2(x_1, x_2)$ that satisfy

$$\Psi_1^*(x_1)\Psi_1(x_1) = \int dx_2 dx_3...dx_N \Psi^*(x_1,..,x_N)\Psi(x_1,..,x_N), \tag{25}$$

$$\Psi_2^*(x_1, x_2)\Psi_2(x_1, x_2) = \int dx_3...dx_N \Psi^*(x_1,..,x_N)\Psi(x_1,..,x_N). \tag{26}$$

Now, we apply the electron-independent approximation assuming the two-body wave function $\Psi_2$ as follow

$$\Psi_2(x_1, x_2) = \Psi_1(x_1 + x_2)\chi(x_1 - x_2) \approx \Psi_1(x_1 + x_2). \tag{27}$$

Then, we obtain

$$\mathcal{H}'[\Psi_1] = -\frac{1}{2m}\frac{Nq^2A_0^2}{m\widetilde{\omega}_c^{(MB)}}\int dx \, \Psi_1^*(x)(-i\nabla)^2\Psi_1(x). \tag{28}$$

Combing this energy term with other terms in the Hamiltonian, we have

$$\langle 0| H_U |0\rangle \approx \sum_{j=1}^{N} \frac{\mathbf{p}_j^2}{2\widetilde{m}^{(MB)}} + V_{\text{lattice}}(\mathbf{r}_j - \frac{1}{2}\xi^2\mathbf{p}_j \times \hat{\mathbf{z}}) + \frac{q^2A_0^2\sigma_z^j}{2m}, \tag{29}$$

$$\widetilde{m}^{(MB)} = m(1 + \frac{Nq^2A_0^2}{m\omega_c}). \tag{30}$$

This Hamiltonian has been derived in Ref. [1]. Following the same steps as we did in the single particle case, the eq. (29) leads to the same form of effective-gauge-field-coupled Hamiltonian eq. (14) with a substitution

$$\begin{aligned}\tilde{\omega}_c &\to \widetilde{\omega}_c^{(MB)}, \\ \tilde{m} &\to \widetilde{m}^{(MB)}.\end{aligned} \tag{31}$$

## II. PHOTON STATE IN THE CHIRAL CAVITY WITH KAGOME LATTICE INSIDE

We start by computing the averaged photon number $\bar{N} = \langle g | a^\dagger a | g \rangle$, where $|g\rangle$ is the ground state of the system in the lab frame

$$
\begin{aligned}
\bar{N} &= \langle g | UU^\dagger a^\dagger a UU^\dagger | g \rangle \\
&= \langle 0 | a^\dagger a + \xi \mathbf{p} \cdot (\boldsymbol{\epsilon}^* a^\dagger + \boldsymbol{\epsilon} a) + \xi^2 (\mathbf{p} \cdot \boldsymbol{\epsilon}^*)(\mathbf{p} \cdot \boldsymbol{\epsilon}) | 0 \rangle \\
&= \langle 0 | \frac{1}{2} \xi^2 \mathbf{p}^2 | 0 \rangle .
\end{aligned}
\tag{32}
$$

This shows that the photon number is proportional to the averaged kinetic energy of the electronic state discussed in the main text.

Next, we compute the transportation of photons through the waveguide coupled to the chiral cavity. As presented in the main text, the Hamiltonian of photons in the cavity and in the waveguide is

$$
H_0 = \omega_c a^\dagger a + \omega_k b_k^\dagger k_k
\tag{33}
$$

where $\omega_c$ is the cavity frequency, and $\omega_k$ is the frequency of photons in the waveguide of wave number $k$. The coupling between the cavity photons and the waveguide photons is described by

$$
H_g = g_k a^\dagger b_k + g_k^* b_k^\dagger a
\tag{34}
$$

In the AD frame, the coupling Hamiltonian becomes:

$$
H_g^{(AD)} = g_k a^\dagger b_k + g_k^* b_k^\dagger a - g_k \xi \mathbf{p} \cdot \boldsymbol{\epsilon} b_k - g_k^* \xi \mathbf{p} \cdot \boldsymbol{\epsilon}^* b_k^\dagger .
\tag{35}
$$

The transmission amplitude can be calculated within the time dependent perturbation:

$$
\begin{aligned}
t_k &= \langle g, k, t = +\infty | \mathcal{T} e^{-i \int dt (H_0 + H_g \delta(t))} | g, k, t = -\infty \rangle \\
&\approx \langle g, k, t = +\infty | e^{-iH_0 t} \left[ 1 - i \int_{-\infty}^{+\infty} dt_1 V(t_1) - \int_{-\infty}^{+\infty} dt_1 \int_{-\infty}^{t_1} dt_2 V(t_1) V(t_2) \right] | g, k, t = -\infty \rangle
\end{aligned}
\tag{36}
$$

where we have defined

$$
V(t) = e^{itH_0} H_g \delta(t) e^{-itH_0} .
\tag{37}
$$

As before, it is easier to work in AD frame, and we compute $t_k$ order by order. The zeroth order is trivially $t_k^{(0)} = 1$. The first order is $t_k^{(1)} = 0$ because of the annihilation operators acting on the zero photon state. The second order term is the first non-trivial term. To compute $t_k^{(2)}$, it would be helpful to first check the commutation relation

$$
[H_0, H_g] = (\omega_c - \omega_k) g_k a^\dagger b_k - (\omega_c - \omega_k) g_k^* a b_k^\dagger
\tag{38}
$$

and hence

$$ad_{H_0}^n := \underbrace{[H_0, [H_0, \cdots [H_0}_{nH_0}, H_g]]] = (\omega_c - \omega_k)g_k a^\dagger b_k - (\omega_c - \omega_k)g_k^* a b_k^\dagger. \tag{39}$$

With this, the Hadamard formula leads to

$$\begin{aligned}
V(t) &= e^{itH_0} H_g \delta(t) e^{-itH_0} \\
&= \sum_{n=0}^{\infty} \frac{(it)^n}{n!} ad_{H_0}^n H_g \delta(t) \\
&= g_k(t) a^\dagger b_k + g_k^*(t) a b_k^\dagger,
\end{aligned} \tag{40}$$

where we have defined

$$g_k(t) = e^{it(\omega_c - \omega_k)} \delta(t) g_k. \tag{41}$$

Now the $t_k^{(2)}$ is ready to be computed

$$\begin{aligned}
t_k^{(2)} &= -\int_{-\infty}^{+\infty} dt_1 \int_{-\infty}^{t_1} dt_2 \, \langle 0, k| \, U^\dagger V(t_1) U U^\dagger V(t_2) U \, |0, k\rangle \\
&= -\int_{-\infty}^{+\infty} dt_1 \int_{-\infty}^{t_1} dt_2 \, \langle 0, k| \, g_k^*(t_1) g_k(t_2) \omega_k \, |0, k\rangle \\
&\quad -\int_{-\infty}^{+\infty} dt_1 \int_{-\infty}^{t_1} dt_2 \, \langle 0, k| \, g_k(t_1) g_k^*(t_2) \xi^2 (\mathbf{p} \cdot \boldsymbol{\epsilon})(\mathbf{p} \cdot \boldsymbol{\epsilon}^*)(1 + \omega_k) \, |0, k\rangle \\
&\quad -\int_{-\infty}^{+\infty} dt_1 \int_{-\infty}^{t_1} dt_2 \, \langle 0, k| \, g_k^*(t_1) g_k(t_2) \xi^2 (\mathbf{p} \cdot \boldsymbol{\epsilon}^*)(\mathbf{p} \cdot \boldsymbol{\epsilon}) \omega_k \, |0, k\rangle \\
&= -\frac{1}{2}\omega_k |g_k|^2 - (\omega_k + \frac{1}{2})|g_k|^2 \langle 0| \, \frac{1}{2}\xi^2 p^2 \, |0\rangle \\
&= -\frac{1}{2}\omega_k |g_k|^2 - (\omega_k + \frac{1}{2})|g_k|^2 \bar{N}
\end{aligned} \tag{42}$$

which, combining with the unity from the zeroth order, reproduces the formula we claimed in the main text.


\end{document}